\documentclass[11pt,a4paper]{article} 
\usepackage{jheppub}

\usepackage{epsfig,multicol,bbm}
\usepackage{graphicx}
\usepackage{amsmath}
\usepackage{bm}
\usepackage{multirow}
\allowdisplaybreaks[1]

\def\sss{\scriptscriptstyle}
\newcommand{\nl}{\nonumber\\}
\newcommand{\bea}{\begin{eqnarray}}
\newcommand{\eea}{\end{eqnarray}}
\def\hp{{h^{\sss +}}}
\def\hm{{h^{\sss -}}}
\newcommand{\kpp}{{k^{\sss ++}}}
\newcommand{\kmm}{{k^{\sss --}}}
\newcommand\fverbdo{\egroup\medskip\noindent%
			\fbox{\unhbox\fverbbox}\ }
\newcommand\fverbit{\egroup\item[\fbox{\unhbox\fverbbox}]}
\newbox\fverbbox

\newcommand{\al}{\alpha}
\newcommand{\ga}{\gamma}
\newcommand{\la}{\lambda}
\newcommand{\si}{\sigma}

\newcommand{\De}{\Delta}
\newcommand{\Ga}{\Gamma}
\newcommand{\Om}{\Omega}

\def\gtsim{\mathrel{\hbox{\raise0.2ex
\hbox{$>$}\kern-0.75em\raise-0.9ex\hbox{$\sim$}}}}
\def\ltsim{\mathrel{\hbox{\raise0.2ex
\hbox{$<$}\kern-0.75em\raise-0.9ex\hbox{$\sim$}}}}

\preprint{KIAS-P12056}

\title{ Can Zee-Babu model implemented with scalar dark matter 
explain both Fermi-LAT 130 GeV $\gamma$-ray excess and neutrino 
physics ? }

\author[a]{Seungwon Baek}
\author[a]{P. Ko,}
\author[a]{Hiroshi Okada,}
\author[a,b]{and Eibun Senaha}

\affiliation[a]{School of Physics, KIAS, Seoul 130-722, Korea}
\affiliation[b]{Department of Physics, Nagoya University, Nagoya 464-8602, Japan}

\emailAdd{swbaek@kias.re.kr}
\emailAdd{pko@kias.re.kr}
\emailAdd{hokada@kias.re.kr}
\emailAdd{senaha@kias.re.kr}

\abstract{
We extend the Zee-Babu model for the neutrino masses and mixings 
by first incorporating a scalar dark matter $X$ with $Z_2$ symmetry and then $X$ and  a dark scalar $\varphi$
with global $U(1)$ symmetry.  
In the latter scenario the singly and doubly charged scalars that are new in the Zee-Babu model 
can explain the large annihilation cross section of a dark matter pair into two 
photons as hinted by the recent analysis of the Fermi $\gamma$-ray space telescope
data. These new scalars can also enhance the 
$B ( H \rightarrow \gamma\gamma )$, as the recent LHC results may suggest. 
The dark matter relic density can be explained. The direct detection rate 
of the dark matter is predicted to be about one order of magnitude 
down from the current experimental bound in the first scenario.
}
\begin{document} 
\maketitle
\section{Introduction\label{sec:intro}}
Although it is well known that the dark matter (DM) constitutes about 27\% of 
the total mass density of the universe, {\it i.e.}
$\Omega_{\rm DM} h^2 = 0.1199 \pm 0.0027$~\cite{Planck:2013}, 
its existence has only been inferred from
the gravitational interaction. And its nature is still unknown. 
If the DM is weakly interacting massive particle
(WIMP), it may reveal itself via non-gravitational interactions, for
example, by pair annihilation into ordinary standard model (SM) particles
including photon~\cite{Zeldovich:1980st}.
In this case, the DM relic abundance is roughly related to
the pair annihilation cross section at freezeout, 
$\langle \sigma v \rangle_{\rm th}$, as
\bea
 \Omega_{\rm DM} h^2 = \frac{3 \times 10^{-27} {\rm cm^3/s}}
{\langle \sigma v \rangle_{\rm th}}.
\label{eq:Om}
\eea
Recently Refs.~\cite{Bringmann:2012,Weniger:2012} claim that the Fermi
$\gamma$-ray space telescope may have seen excess of the photons with 
$E_\gamma \sim 130$ GeV from the center of the Milky Way compared 
with the background. 
Interpreting its origin as the annihilation of a pair of DM particles, 
they could obtain the annihilation cross section to be about 4\% of that at freezeout:
\bea
 \langle \sigma v \rangle_{\gamma\gamma} \approx 0.04  \langle \sigma v \rangle_{\rm th}
\approx 0.04 {\rm ~pb} \approx 1.2 \times 10^{-27} {\rm cm^3/s}.
\label{eq:sigmav_Fermi}
\eea
Since DM is electrically neutral, the pair annihilation process
into photons occurs through loop-induced diagrams. Naively we expect
\bea
\frac{ \langle \sigma v \rangle_{\gamma\gamma}}{ \langle \sigma v \rangle_{\rm th}} = 
\left(\frac{\alpha_{\rm em}}{\pi} \right)^2 \sim 10^{-5}.
\eea
So the observed value in (\ref{eq:sigmav_Fermi}) is rather large,  and 
we may need new electrically charged particles running inside the loop beyond the SM.  
Many new physics scenarios were speculated within various CDM models by 
this observation ~\cite{recent}. 

The so-called `Zee-Babu model'~\cite{Zee,Babu,Babu:2003} provides new 
charged scalars, $h^+, k^{++}$ at electroweak scale, in addition to the SM particles. 
These new charged scalars carry two units of lepton number and can generate
Majorana neutrino masses via two-loop diagrams. The diagrams are finite
and calculable. The neutrino masses are naturally small without the need to 
introduce the right-handed neutrinos for seesaw mechanism. One of the neutrinos
is predicted to be massless in this model. Both normal and inverse hierarchical
pattern of neutrino masses are allowed. The observed
mixing pattern can also be accommodated. The model parameters are strongly
constrained by the neutrino mass and mixing data, the radiative muon decay,
$\mu \to e \gamma$, and $\tau \to 3 \mu$ decay~\cite{Hirsch}.

It would be very interesting to see if the new charged particles in the Zee-Babu
model can participate in some other processes in a sector independent of neutrinos.
In the first part of this paper, we minimally extend the Zee-Babu model to incorporate the DM.
In the later part we will consider more extended model with global $U(1)_{B-L}$ symmetry and
an additional scalar which breaks the global symmetry~\cite{Lindner:2011it}.

In the first scenario, we introduce a real scalar dark matter $X$ with a discrete $Z_2$ symmetry under 
which the dark matter transforms as $X \to -X$ in order to guarantee its stability.
The renormalizable interactions between the scalar DM $X$ with the Higgs field 
and the Zee-Babu scalar fields provide a Higgs portal between the SM sector 
and the DM.
We show that the Zee-Babu scalars and their interactions with the DM particle
can explain the DM relic density.
The branching ratio of Higgs to two photons, $B(H\to \gamma\gamma)$, can
also be enhanced as implied by the recent LHC results~\cite{CMS}.
The spin-independent cross section of the dark matter scattering off the
proton, $\sigma_p$, is less than about $1 \times 10^{-9}$ pb, which can be
probed at next generation searches. Although the charged Zee-Babu scalars enhance
the $X X \to \gamma \gamma$ process, it turns out the the current experimental constraints
on their masses do not allow the annihilation cross section to reach (\ref{eq:sigmav_Fermi}).

In the extended scenario, we consider a complex scalar dark matter $X$ and a dark scalar 
$\varphi$~\cite{Lindner:2011it}. The global $U(1)$ symmetry of the original Lagrangian is broken down
to $Z_2$ by $\varphi$ getting a vacuum expectation value (vev). 
We show that both the dark matter relic abundance and the Fermi-LAT gamma-ray line signal
can be accommodated via two mechanisms.

This paper is organized as follows.  
In Section~\ref{sec:Z2_model}, we define our model by including the scalar 
DM in the Zee-Babu model, and consider theoretical constraints on the 
scalar potential. 
Then we study various DM phenomenology.
We calculate the dark matter relic density and  the annihilation cross 
section $\langle \sigma v \rangle_{\gamma \gamma}$ in our
model. We also predict the cross section for the DM and proton scattering
and the branching ratio for the Higgs decay into two photons, 
$B(H \rightarrow \gamma \gamma)$. And we consider the implication
on the neutrino sector.
In Section~\ref{sec:u1}, we consider the DM phenomenology
in the extended model.
 We conclude in Section~\ref{sec:Conclusions}.

\section{The $Z_2$ model\label{sec:Z2_model}}
We implement the Zee-Babu model for radiative generation of neutrino masses and mixings, 
by including a real scalar DM $X$ with $Z_2$ symmetry $X \rightarrow -X$. 
All the possible renormalizable interactions involving the scalar fields are given by
\begin{eqnarray}
{\cal L} & = & {\cal L}_{\rm Babu} + {\cal L}_{\rm Higgs + DM}
\\
{\cal L}_{\rm Babu} & = & f_{ab} l_{aL}^{T i} C l_{bL}^j \epsilon_{ij} h^+ 
+ h_{ab}^{'} l_{aR}^{T} C l_{bR} k^{++} + h.c. 
\label{eq:yuk}
\\
-{\cal L}_{\rm Higgs + DM} & = & 
-\mu_H^2 H^\dagger H + \frac{1}{2} \mu_X^2 X^2 + \mu_h^2 h^+ h^- +\mu_k^2 \kpp \kmm \nl 
&& + (\mu_{hk} \hm \hm \kpp + h.c.)\nonumber\\
&& + \lambda_H (H^\dagger H)^2 + \frac{1}{4}\lambda_X X^4
+\lambda_h (h^+ h^-)^2 + \lambda_k (\kpp \kmm)^2 \nonumber\\
&+& \frac{1}{2} \lambda_{HX} H^\dagger H X^2 + \lambda_{Hh} H^\dagger H h^+ h^- + \lambda_{Hk} H^\dagger H \kpp \kmm \nonumber\\
&+& \frac{1}{2}\lambda_{Xh} X^2 h^+ h^-  + \frac{1}{2}\lambda_{Xk} X^2 \kpp \kmm
 + \lambda_{hk} h^+ h^- \kpp \kmm.
\label{eq:pot}
\end{eqnarray}
Note that our model is similar to the model proposed by J. Cline~\cite{Cline:2012}.
However we included the interaction between the new charged scalar and the SM leptons
that are allowed by gauge symmetry, and thus the new charged scalar bosons are not 
stable and cause no problem. 

The original Zee-Babu model was focused on the neutrino physics, and the operators 
of Higgs portal types were not discussed properly. 
It is clear that those Higgs portal operators we include in the 2nd line of (\ref{eq:pot}) 
can enhance $H\rightarrow \gamma\gamma$,  without touching any other decay rates 
of the SM Higgs boson, as long as $h^\pm$  and $k^{\pm\pm}$ are heavy enough that 
the SM Higgs decays into these new scalar bosons are kinematically forbidden.

\subsection{Constraints on the potential}
We require $\mu_X^2$, $\mu_h^2$ and $\mu_k^2$ to be positive. Otherwise 
the imposed $Z_2$ symmetry $X \to -X$ or the electromagnetic U(1) symmetry
could be spontaneously broken down.  
Since the masses of $X$, $\kpp$ and $\hp$ have contributions from the electroweak 
symmetry breaking as
\bea
  m_X^2 &=& \mu_X^2 + \frac{1}{2} \lambda_{HX} v_H^2, \nl
  m_\hp^2 &=& \mu_h^2 + \frac{1}{2} \lambda_{Hh} v_H^2, \nl
  m_\kpp^2 &=& \mu_k^2 + \frac{1}{2} \lambda_{Hk} v_H^2,
\eea
we obtain the conditions on the quartic couplings
\bea
\lambda_{HX} < \frac{2 m_X^2}{v_H^2}, \quad
\lambda_{Hh} < \frac{2 m_\hp^2}{v_H^2}, \quad
\lambda_{Hk} < \frac{2 m_\kpp^2}{v_H^2}.
\label{eq:sym_br}
\eea
We note that the above conditions are automatically satisfied if
the couplings takes negative values. 
In such a case, however, 
we also need to worry about the behavior of the Higgs potential
for large field values. For example, if we consider only the 
neutral Higgs field, $H$\footnote{We use the same notation with the
Higgs doublet.}, and the dark matter field, $X$, we get
\bea
 V &\sim& \frac{1}{4} \lambda_H H^4 + \frac{1}{4} \lambda_X X^4 + \frac{1}{4}
   \lambda_{HX} H^2 X^2, \nl
   &\sim& \frac{1}{4} \left(\begin{array}{cc} H^2 & X^2 \end{array} \right)
\left(\begin{array}{cc}  \lambda_H & \frac{1}{2} \lambda_{HX} \\
 \frac{1}{2} \lambda_{HX} & \lambda_X\\
 \end{array} \right)
\left(\begin{array}{c}  H^2 \\ X^2
 \end{array} \right),
\label{eq:potential_large}
\eea
for large field values of $H$ and $X$. If the potential is to be
bounded from below, every eigenvalue of the square matrix of the
couplings in (\ref{eq:potential_large}) should be positive, whose
condition is
\bea
 |\lambda_{HX}| < \sqrt{4 \lambda_H \lambda_X}.
\label{eq:ufb}
\eea
This means that even if $\lambda_{HX}$ is negative, its absolute value
should not be arbitrarily large because $\lambda_H=m_H^2/2 v_H^2 \approx 0.13$ 
($m_H \approx 125$ GeV) and $\lambda_X$ is bounded from above so as not to generate
the Landau pole. 
For example, the renormalization group running equations (RGEs) of $\lambda_H$,
$\lambda_X$ and $\lambda_{HX}$ are
given by
\bea
 \frac{d \lambda_{H}}{d \log Q}
&=& \frac{1}{16 \pi^2} \Big(24 \lambda_H^2 + \frac{1}{2} \lambda_{HX}^2 \Big) + \cdots, \nl
 \frac{d \lambda_{X}}{d \log Q}
&=& \frac{1}{16 \pi^2} \Big(18 \lambda_X^2 + 2 \lambda_{HX}^2 \Big) + \cdots, \nl
 \frac{d \lambda_{HX}}{d \log Q}
&=& \frac{\lambda_{HX}}{8 \pi^2} \Big(6 \lambda_H + 3 \lambda_X \Big) + \cdots,
\label{eq:RGE}
\eea
where the dots represents other contributions which are not important
in the discussion. The complete forms of the $\beta$-functions of the quartic couplings are listed
in Appendix \ref{app:RGE}.

The approximate solution for $\lambda_X$ in (\ref{eq:RGE}) 
shows that the Landau pole is generated at the scale
$Q = Q_{\rm EW} \exp\left(1/\beta_H \lambda_X(Q_{\rm EW})\right)$ 
($\beta_H =18/16 \pi^2$). 
If we take the electroweak scale value of the Higgs quartic coupling
to be $\lambda_X(Q_{\rm EW}) \sim 5$, the cut-off scale should be around 1 TeV.
The general condition for the bounded-from-below potential for large
field values is that all the eigenvalues of the matrix
\bea
\left( \begin{array}{cccc}
\lambda_H & \frac{1}{2} \lambda_{HX} & \lambda_{Hh} & \lambda_{Hk} \\
 \frac{1}{2} \lambda_{HX} & \lambda_X & \lambda_{Xh} & \lambda_{Xk} \\
 \lambda_{Hh} & \lambda_{Xh} & 4 \lambda_{h} & 2 \lambda_{hk} \\
 \lambda_{Hk} & \lambda_{Xh} & 2 \lambda_{hk} & 4 \lambda_{k} 
 \end{array} \right)
\eea
should be positive. 

\begin{figure}
\center
\includegraphics[width=8cm]{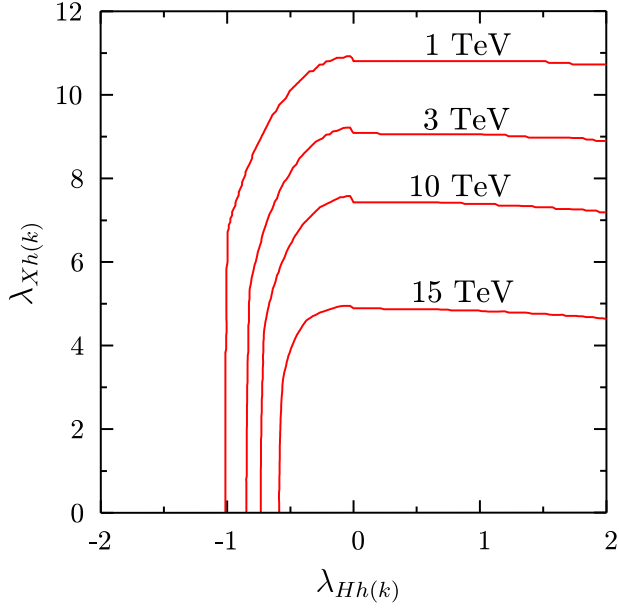}
\caption{The perturbativity bounds, $\lambda_i(Q)<4\pi$ are shown. The each curve denotes $Q=1$, 3, 10 and 15 TeV from top to bottom. We take $\lambda_{Hh}=\lambda_{Hk}$, $\lambda_{Xh}=\lambda_{Xk}$, $\lambda_{hk}=\lambda_{HX}=0$ and $\lambda_X=\lambda_H(\simeq 0.13)$. For the negative $\lambda_{Hk}$, we set $\lambda_h=\lambda_{Hh}^2/(2\lambda_H)$ and $\lambda_k=\lambda_{Hk}^2/(2\lambda_H)$ while $\lambda_h=\lambda_k=\lambda_H$ for positive $\lambda_{Hk}$. }
\label{fig:RG}
\end{figure}

In the following discussion, we require that all scalar quartic couplings 
($\lambda_i$) be perturbative up to some scale $Q$. 
To this end, we solve the one-loop RGEs of those quartic couplings given 
in Appendix \ref{app:RGE}.  For the moment, we do not include new Yukawa 
couplings defined in Eq.~(\ref{eq:yuk}), and we adopt the criterion 
$\lambda_i(Q)<4\pi$ in this analysis.
In Fig.~\ref{fig:RG}, the perturbativity bounds are shown in the 
$\lambda_{Xh(k)}$-$\lambda_{Hh(k)}$ plane. We take $Q=1$, 3, 10 and 15 TeV,
which are denoted by the red curves from top to bottom.
For other parameters, we fix $\lambda_{Hh}=\lambda_{Hk}$, 
$\lambda_{Xh}=\lambda_{Xk}$, $\lambda_{hk}=\lambda_{HX}=0$ and 
$\lambda_X=\lambda_H(\simeq 0.13)$. 
As explained above, a certain negative value of $\lambda_{Hk(h)}$ 
may cause the instability of the Higgs potential. 
To avoid this, we set $\lambda_h=\lambda_{Hh}^2/(2\lambda_H)$ and $\lambda_k=\lambda_{Hk}^2/(2\lambda_H)$ for $\lambda_{Hk(h)} <0$.
For $\lambda_{Hk(h)}>0$, on the other hand, 
$\lambda_h=\lambda_k=\lambda_H$ is taken.  As we see from the plot, 
$\lambda_{Xk(h)}\simeq 7-11$ is possible if $Q=1$ TeV.

The theoretical arguments (\ref{eq:sym_br}) and (\ref{eq:ufb}) 
restrict $\lambda_{HX}$ to lie roughly to the range, $(-1.6, 0.6)$.
Similarly, we have $\lambda_{Hh(k)}\ltsim 0.7$ for $m_{h^+(k^{++})}=150$ GeV.

%


\subsection{$X X \to \gamma \gamma$ and Fermi-LAT 130 GeV $\gamma$-ray excess}
The annihilation cross section for $X X \to \gamma \gamma$ is given by
\bea
\langle \sigma v \rangle_{\gamma\gamma} = \frac{\sum |{\cal M}|^2}{64 \pi m_X^2},
\eea
where the amplitude-squared summed over the photon polarization is
\bea
\sum |{\cal M}|^2 &=& \frac{\alpha_{\rm em}^2}{2 \pi^2} \Bigg|
\lambda_{Xh}  A_0(\tau_{\hp})
+4 \lambda_{Xk}  A_0(\tau_{\kpp}) \nl
&+& \frac{\lambda_{HX} v_H^2}{s-m_H^2+i m_H \Ga_H}
\bigg[\frac{g^2}{2 \tau_W}  \Big( Q_t^2 N_C A_{1/2}(\tau_t)+A_1(\tau_W)\Big) \nl
&+&\lambda_{Hh}  A_0(\tau_{\hp})
+4 \lambda_{Hk}  A_0(\tau_{\kpp})
\bigg] \Bigg|^2,
\label{eq:XX2rr}
\eea
with $\tau_i = 4 m_i^2/s (i=\hp,\kpp,t,W)$.
The loop functions are 
\bea
A_0(\tau) &=& 1-\tau f(\tau), \nl
A_{1/2}(\tau) &=& -2 \tau\Big[1+(1-\tau)f(\tau)\Big], \nl
A_1(\tau) &=& 2+3\tau+3\tau(2-\tau)f(\tau),
\eea
where
\bea
f(\tau) = \left\{
\begin{array}{l}
\arcsin^{2}\sqrt{1/\tau}, \quad (\tau \geq 1) \\
-\frac{1}{4}\Big[\log\frac{1+\sqrt{1-\tau}}{1-\sqrt{1-\tau}}-i \pi \Big]^2,
\quad(\tau < 1).
\end{array}
\right.
\eea

\begin{figure}
\centering
\includegraphics[width=0.45\textwidth]{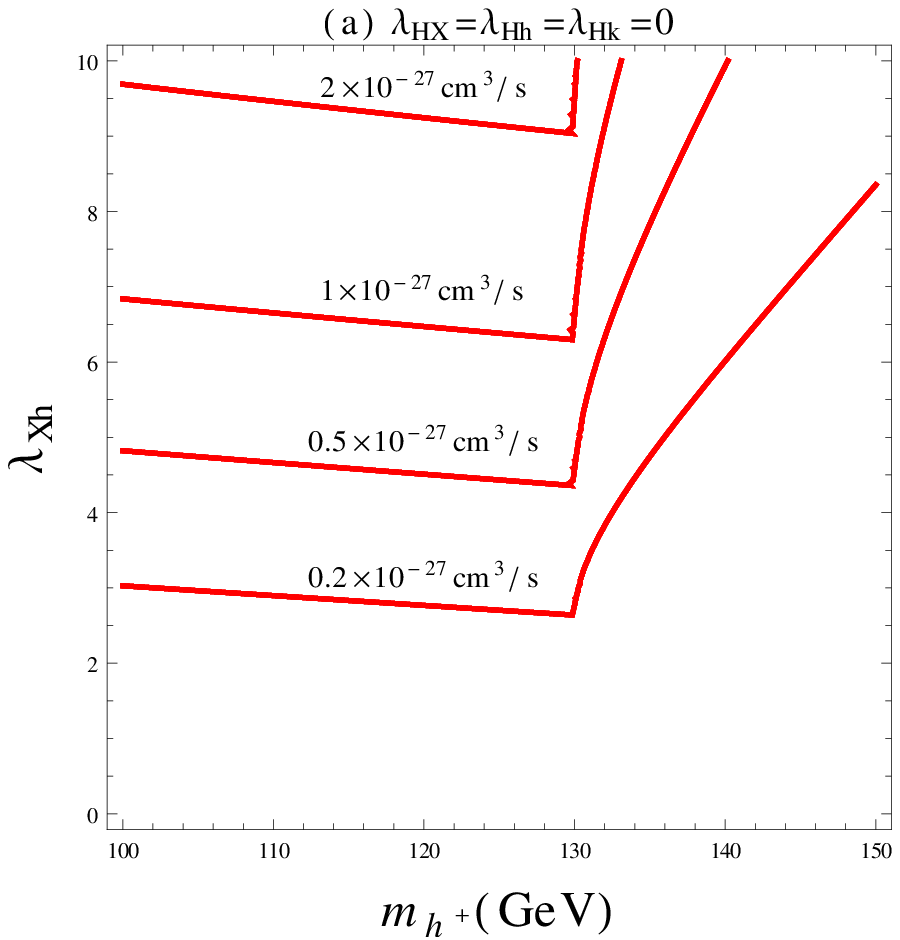}
\includegraphics[width=0.45\textwidth]{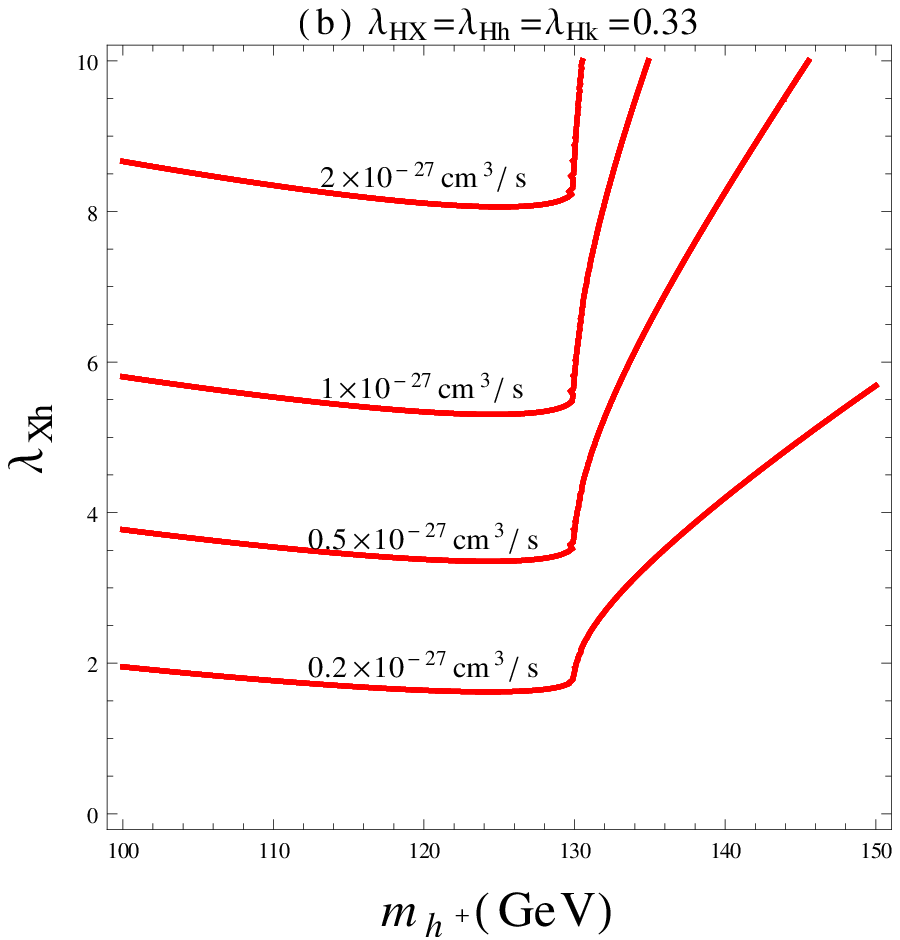}
\caption{Contour plot of 
$\langle \sigma v \rangle_{\gamma\gamma}=(2, 1, 0.5, 0.2) \times 10^{-27} {\rm cm^3/s}$
(from above) in $(m_{\hp},\lambda_{Xh})$ plane. 
We set $m_X =130~{\rm GeV}$, $m_H=125 ~{\rm GeV}$, $m_k=500 ~{\rm GeV}$
and $\lambda_{Xk}=5, \lambda_{HX}=\lambda_{Hh}=\lambda_{Hk}=0 (0.33)$ in the
left (right) panel.
}
\label{fig:XX2rr}
\end{figure}
Although the contribution of the doubly-charged Higgs $\kpp$ to 
$\langle \sigma v \rangle_{\gamma\gamma}$ is $2^4=16$ times larger than that of the 
singly-charged Higgs $\hp$ when their masses are similar to each other, 
this option is ruled out by the recent LHC searches for the doubly-charged 
Higgs boson~\cite{CMS:Hpp}. Depending on the decay channels, the 95\% CL lower
limit on the mass of the doubly-charged Higgs boson is in the range, 204--459 GeV.
To be conservative, we set $m_\kpp = 500$ GeV. 
In Figure~\ref{fig:XX2rr}, we show a contour plot for the
annihilation cross section into two photons: 
$\langle \sigma v \rangle_{\gamma\gamma} \approx (2, 1, 0.5, 0.2) 
\times 10^{-27} {\rm cm^3/s}$ (from above) in the $(m_{\hp},\lambda_{Xh})$ plane.
We set $m_X =130~{\rm GeV}$, $m_H=125 ~{\rm GeV}$, $m_\kpp=500 ~{\rm GeV}$
and $\lambda_{Xk}=5, \lambda_{HX}=\lambda_{Hh}=\lambda_{Hk}=0 ~(0.33)$ 
in the left (right) panel. 
We can see that by turning on the process, $XX \rightarrow H \rightarrow \gamma\gamma$,
with $\lambda_{HX} = 0.33$ (right panel), we can reduce $\lambda_{Xh}$ to get
$\langle \sigma v \rangle_{\gamma\gamma} =1 \times 10^{-27} \, {\rm cm^3/s}$ to explain the Fermi-LAT gamma-ray line signal, but
not significantly enough to push the cut-off scale much higher than the electroweak scale.
As we will see in the following section, the $\langle \sigma v \rangle_{\gamma\gamma} =1 \times 10^{-27} \, {\rm cm^3/s}$
is not consistent with the current DM relic abundance.

%
\subsection{Thermal relic density and direct detection rate}
%

%
\begin{figure}
\centering
\includegraphics[width=0.5\textwidth]{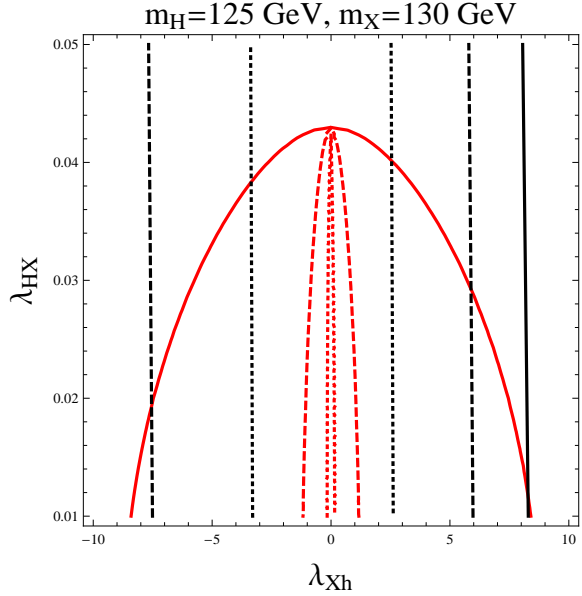}
\caption{
The contour plot of $\Omega_{\rm DM} h^2=0.1199$ (red lines) and
$\langle \sigma v \rangle_{\gamma\gamma} = 0.2 \times 10^{-27} {\rm cm^3/s}$ 
(black lines) in the ($\lambda_{Xh}$,$\lambda_{HX}$) plane for the choices
$m_\hp=$ 150, 140, 130 GeV (solid, dashed, dotted lines).
For other parameters we set $m_X = 130$ GeV, $m_H= 125$ GeV, $m_k = 500$ GeV, 
$\lambda_{Xk}=5$, $\lambda_{Hh}=\lambda_{Hk}=0.5$. 
}
\label{fig:relic}
\end{figure}

Contrary to J. Cline's model~\cite{Cline:2012}, the DM relic density 
in our model is not necessarily
correlated with the $\langle \sigma v \rangle_{\gamma\gamma}$, since it
is mainly determined by $\lambda_{HX}$ for relatively heavy scalars 
($\gtrsim 150$ GeV). In this case the main DM annihilation channels are
$X X \to H \to$ SM particles, where the SM particles are
$W^+ W^-$, $Z Z$, $b \bar{b}$, {\it etc}.
As  $m_{\hp (\kpp)}$ becomes comparable with $m_X$, the 
$X X \to \hp \hm  (\kpp \kmm)$ modes can open, even in cases
$m_X < m_\hp (m_\kpp)$ due to the kinetic energy of $X$ at freeze-out time. 
This can be seen in Figure~\ref{fig:relic}, where we show the contour plot of 
$\Omega_{\rm DM} h^2=0.1199$ (red  lines) in the ($\lambda_{Xh}$,$\lambda_{HX}$) 
plane for the choices $m_\hp=$ 150, 140, 130 GeV 
(shown in solid, dashed, dotted lines respectively).
We fixed other parameters to be $m_X = 130$ GeV, $m_H= 125$ GeV, $m_k = 500$ GeV, 
$\lambda_{Xk}=5$, $\lambda_{Hh}=\lambda_{Hk}=0.5$. For $m_\hp =130$ GeV, the 
annihilation mode $XX \rightarrow \hp \hm$ dominates even for very small coupling
$\lambda_{Xh}$ (the red dotted line).
The black vertical lines are the constant contour lines of 
$\langle \sigma v \rangle_{\gamma\gamma} = 0.2 \times 10^{-27} {\rm cm^3/s}$.
We can see that the maximum value for the Fermi-LAT gamma-ray line signal
which is consistent with the relic density is $\langle \sigma v \rangle_{\gamma\gamma} 
= 0.2 \times 10^{-27} {\rm cm^3/s}$ when $m_{h^+}=150$ GeV. This cross section
is smaller than the required value in (\ref{eq:sigmav_Fermi}) by factor 6.


%
\begin{figure}
\centering
\includegraphics[width=0.7\textwidth]{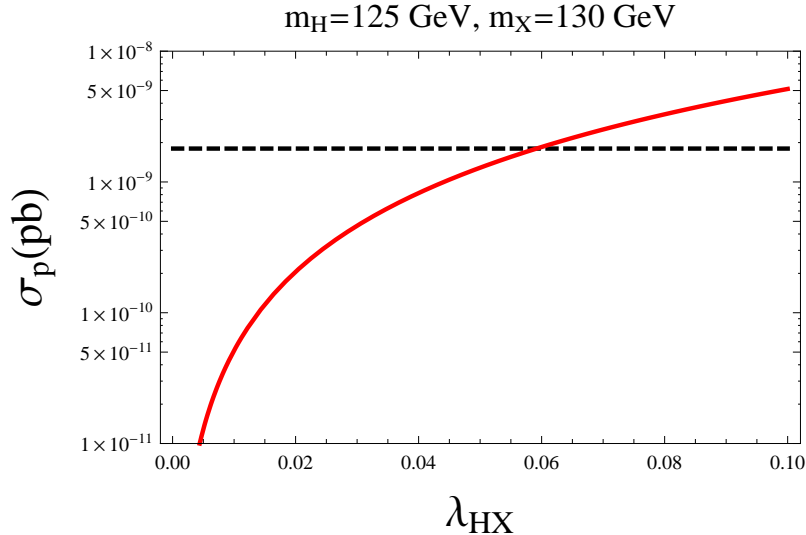}
\caption{The spin-independent cross section of dark matter scattering off proton, 
$\sigma_p$,  as a function of $\lambda_{HX}$ (red solid line) and 
$\sigma_p=1.8 \times 10^{-9}$ pb line above which is excluded by LUX
(black dashed line). We take  $m_H = 125$ GeV and $m_X = 130$ GeV.
}
\label{fig:dd}
\end{figure}
Figure~\ref{fig:dd} shows the cross section of dark matter scattering off proton, $\sigma_p$, 
as a function of $\lambda_{HX}$ (red solid line) and $\sigma_p=1.8 \times 10^{-9}$ pb line
(black dashed line)
above which is excluded by LUX~\cite{LUX:2013} at 90\% C.L. This cross section
is determined basically only by $\lambda_{HX}$ at tree level by the SM Higgs exchange, 
when we fix $m_X=130$ GeV.
We can see that $\lambda_{HX} \lesssim 0.06$ to satisfy the LUX upper bound.

\subsection{$H\rightarrow \gamma\gamma$}
\label{sec:Hrr}

In this scenario the decay width of $H \to \gamma \gamma$ is modified, whereas 
other Higgs decay widths are intact:  
\bea
\Gamma(H \to \gamma\gamma) &=& \frac{ \alpha_{\rm em}^2 v_H^2}{64 \pi^3 m_H} \Bigg|
\la_{Hh} A_0(\tau_{\hp}) +4 \la_{Hk} A_0(\tau_{\kpp}) \nl
&+& {g^2 \over 2 \tau_W}\Big[A_1(\tau_W) + \sum_{f=t,b} Q_f^2 N^f_c A_{1/2}(\tau_f)\Big]
\Bigg|^2,
\eea
where $\tau_i = 4 m_i^2/m_H^2\, (i=f, W, \hp, \kpp)$.

\begin{figure}
\centering
\includegraphics[width=0.45\textwidth]{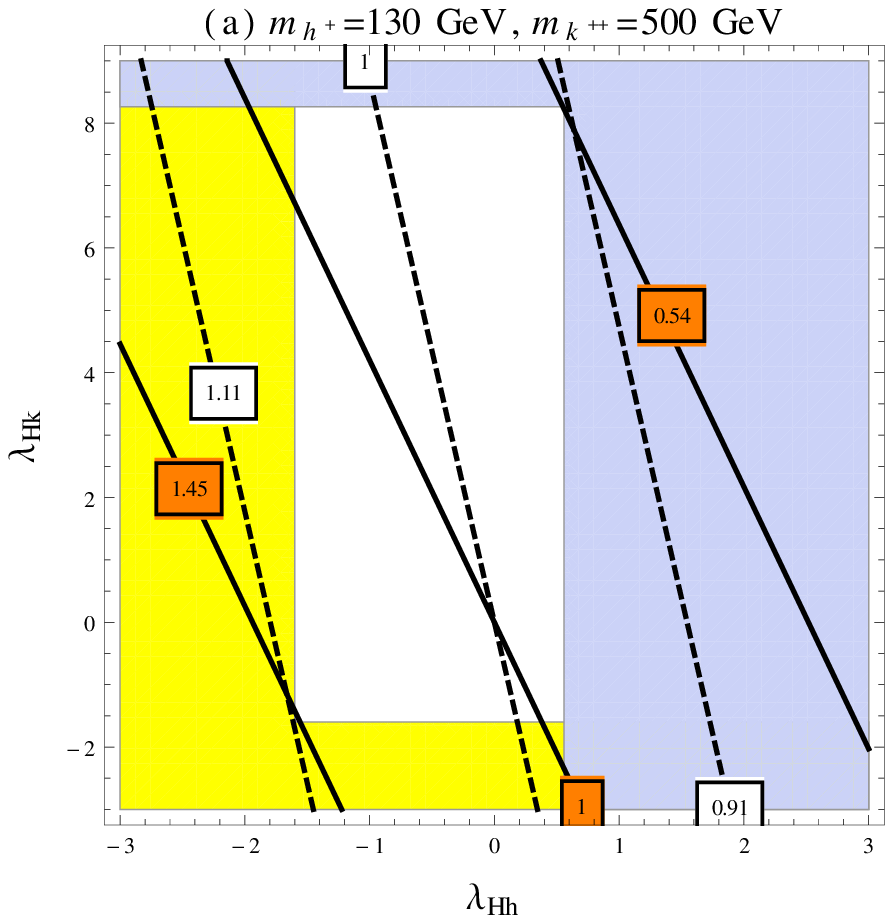}
\includegraphics[width=0.45\textwidth]{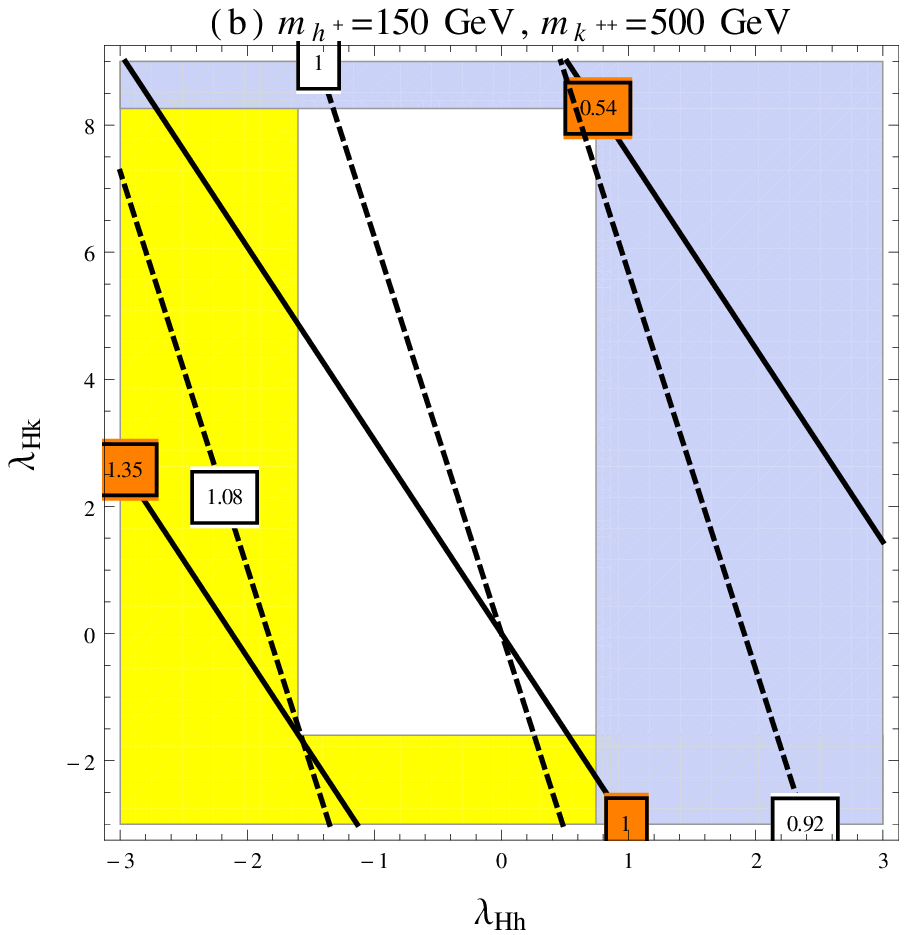}
\caption{A contour plot for constant
$\Gamma(H \to \gamma\gamma)/\Gamma(H \to \gamma\gamma)^{\rm SM}$ (black solid lines) and
$\Gamma(H \to Z \gamma)/\Gamma(H \to Z\gamma)^{\rm SM}$ (black dashed lines) in the 
$(\lambda_{Hh},\lambda_{Hk})$ plane. 
The shaded regions are disfavored by (\ref{eq:sym_br}) (blue)
and by (\ref{eq:ufb}) (yellow).
We set $m_\hp = 130~ (150)$ GeV for the left (right) panel and fixed  $m_\kpp = 500$ GeV.
}
\label{fig:h2rr}
\end{figure}
In Figure~\ref{fig:h2rr}, we show contour plots for constant
$\Gamma(H \to \gamma\gamma)/\Gamma(H \to \gamma\gamma)^{\rm SM}$ (black solid lines) and
$\Gamma(H \to Z \gamma)/\Gamma(H \to Z\gamma)^{\rm SM}$ (black dashed lines) in the 
$(\lambda_{Hh},\lambda_{Hk})$ plane. 
For this plot we set $m_\hp = 130 ~(150)$ GeV for the left (right) panel
and fixed  $m_\kpp = 500$ GeV.  The shaded regions are disfavored by 
(\ref{eq:sym_br}) (blue) and by (\ref{eq:ufb}) (yellow).
The ratios depend basically only on the coupling constants $\lambda_{Hh}$ and $\lambda_{Hk}$
as well as the masses $m_\hp$ and $m_\kpp$. And the ratios are not necessarily 
correlated with the $\langle \sigma v\rangle_{\gamma \gamma}$
which are controlled by $\lambda_{Xh}$ and $\lambda_{Xk}$.
We can conclude that 
\begin{eqnarray*}
0.54 \lesssim & \Gamma(H \to \gamma\gamma)/\Gamma(H \to \gamma\gamma)^{\rm SM} 
& \lesssim 1.45 ~
(1.35)
\\
0.91 \lesssim & \Gamma(H \to Z\gamma)/\Gamma(H \to Z\gamma)^{\rm SM} 
& \lesssim 1.11 ~
(1.08)
\end{eqnarray*}
for the left (right) panel. That is, the $H \to \gamma\gamma$ channel can
be enhanced (reduced) significantly, whereas the $H \to Z\gamma$ channel can change
only upto $\sim 10 \%$.

\subsection{Implications for neutrino physics}

So far, we did not consider the constraints from neutrino sector and charged lepton
flavor violation. In fact, these constraints are rather severe, if we assume that the 
observed neutrino masses and mixings are entirely from the Zee-Babu mechanism.
One cannot afford light $h^\pm$ or $k^{\pm\pm}$, because of the constraints from 
charged LFV: $m_\hp \gtrsim 240$ GeV from 
$B(\mu \rightarrow e \gamma) < 2.4 \times 10^{-12}$~\cite{PDG} 
and $m_\kpp \gtrsim 770$ GeV from the upper bound on $\tau \rightarrow 3 \mu$ 
decay~\cite{Hirsch}.

If we insist that the Fermi-LAT excess is due to the light $h^\pm$ loop, then 
we get $\mu_{hk} \gtrsim 14$ TeV, which is inconsistent with the constraint
$\mu_{hk} \lesssim 450$ GeV from the vacuum stability bound~\cite{Babu:2003}. 
And there should be additional contributions to neutrino masses, such as from 
dim-5 Weinberg operators.  
If these dim-5 operators are induced through Type-I seesaw mechanism, 
the new physics would not affect our conclusion. 
On the other hand, if the dim-5 operators are induced by TeV scale seesaw,
then the new physics from TeV scale seesaw might affect our conclusions.

Although it is not very satisfactory that the original Zee-Babu model with scalar dark
matter cannot explain both the Fermi-LAT 130 GeV $\gamma$ ray excess and 
neutrino physics simultaneously, it would be more natural to consider the Zee-Babu 
model as a low energy effective theory.  Then it would be natural there could be 
new contributions to the neutrino masses and mixings from dim-5 operators.
The only relevant question would be whether those new physics would affect the 
Fermi-LAT $\gamma$-ray or not. If the new physics is Type-I seesaw, there would
be no new charged particles so that our conclusion would remain valid. 

\section{Spontaneously broken $U(1)_{B-L}$ model}
\label{sec:u1}

As we have seen in the previous section, the simplest extension of Zee-Babu model
to incorporate dark matter with $Z_2$ symmetry, although very predictive, has difficulty in fully explaining the
Fermi-LAT gamma-line anomaly. In this section we consider a next minimal model
where we may solve the problem. We further extend the model by introducing
 $U(1)_{B-L}$ symmetry and
additional complex scalar $\varphi$ to break the global symmetry~\cite{Chang:1988aa,Lindner:2011it}. 

Then the model Lagrangian (\ref{eq:pot}) is modified as
\bea
-{\cal L}_{\rm Higgs + DM} & = & 
-\mu_H^2 H^\dagger H + \mu_X^2 X^* X+\mu_h^2 h^+ h^- + \mu_k^2 k^{++} k^{--}-\mu_\varphi^2 \varphi^* \varphi  \nl
&& + (\mu_{\varphi X} \varphi X X  + h.c.) \nl
&& + (\lambda_\mu \varphi h^- h^- k^{++}   + h.c.) \nl
&& + \lambda_H (H^\dagger H)^2 
 + \lambda_\varphi (\varphi^* \varphi)^2  + \lambda_X (X^* X)^2
 + \lambda_h (h^+ h^-)^2   + \lambda_k (k^{++} k^{--})^2 
 \nl
&& +\lambda_{H\varphi} H^\dagger H \varphi^* \varphi
+\lambda_{HX} H^\dagger H X^* X 
+\lambda_{Hh} H^\dagger H h^+ h^- +\lambda_{Hk} H^\dagger H k^{++}k^{--}  
\nl
&& +\lambda_{\varphi X} \varphi^* \varphi X^* X
+\lambda_{\varphi h} \varphi^* \varphi h^+ h^-
+\lambda_{\varphi k} \varphi^* \varphi k^{++} k^{--} \nl
&& +\lambda_{Xh}  X^* X h^+ h^-  +\lambda_{Xk} X^* X k^{++} k^{--} +\lambda_{hk} h^+ h^- k^{++}k^{--},
\label{eq:pot2}
\eea
where we also replaced the real scalar dark matter $X$ in (\ref{eq:pot}) with the complex scalar field.
The charge assignments of scalar fields are given as follows:
\begin{center}
\begin{tabular}{cccccc}
\hline\hline
                   &   $H$                &  $h^+$  &  $k^{++}$    &    $\varphi$   & $X$ \\
\hline
$U(1)_Y$     &    ${1 \over 2} $ &   $1$     &  $2$           &     $0$           & $0$ \\
\hline
$U(1)_{B-L}$ &    $0 $               &   $2$   &  $2$         &     $2$          & $-1$ \\
\hline
\end{tabular}
\end{center}

We note that the soft lepton number breaking term $\mu_{hk} h^+ h^+ k^{--}$ in (\ref{eq:pot}) 
is now replaced by the $B-L$ preserving $\lambda_\mu \varphi  h^+ h^+ k^{--}$ term. The $U(1)_{B-L}$
symmetry is spontaneously broken after $\varphi$ obtains vacuum expectation value (vev). 
In~\cite{Babu:2003}, it was shown that $\mu_{hk} < {\cal O}(1)\, m_{h^+}$ to make the scalar potential
stable. In this $U(1)_{B-L}$ model, this can be always guaranteed by taking small $\lambda_\mu$, since
$\mu_{hk} = \lambda_\mu v_\varphi$ even for very large $v_\varphi$.
The term
$\mu_{X\varphi} XX \varphi $ leaves $Z_2$ symmetry unbroken after $U(1)_{B-L}$ symmetry breaking. 
Under the remnant $Z_2$ symmetry, $X$ is odd while all others are even.
It appears that the theory is reduced to $Z_2$ model in (\ref{eq:pot}) when $\varphi$ is decoupled from the theory.
But we will see that this is not the case and the effect of $\varphi$ is not easily decoupled.

After $H$ and $\varphi$ fields get vev's, in the unitary gauge we can write 
\bea
 H = \left(\begin{array}{c} 0 \\ {1 \over \sqrt{2}}(v_H + h) \end{array}\right), \quad
\varphi = {1 \over \sqrt{2}} (v_\varphi + \phi) e^{ i \alpha/ v_\varphi},
\eea
where $\alpha$ is the Goldstone boson associated with the spontaneous breaking of
global $U(1)_{B-L}$.  For convenience we also rotated the field $X$ 
\bea
 X \to X e^{-i \alpha/ 2 v_\varphi}
\eea
so that the Goldstone boson does not appear in the $\mu_{\varphi X} \varphi X X$ term. 
Then the Goldstone boson interacts with $X$ via the usual derivative coupling coming
from the kinetic term of $X$-field.

The neutral scalar fields $h$ and $\phi$ can mix with
each other to give the mass eigenstates $H_i (i=1,2)$ by rotating
\bea
\left(\begin{array}{c}  h   \\  \phi \end{array}\right) =
\left(\begin{array}{cc}  c_H & s_H   \\  -s_H  &  c_H \end{array}\right)
\left(\begin{array}{c}  H_1   \\  H_2 \end{array}\right),
\eea
where $c_H \equiv \cos \alpha_H$, $s_H \equiv \sin \alpha_H$, with $\alpha_H$ mixing angle,
and we take $H_{1}$ as the SM-like ``Higgs'' field.
Then mass matrix can be written in terms of mass eigenvalues $m_i^2$ of $H_i (i=1,2)$:
\bea
\left(\begin{array}{cc} 2 \lambda_H v_H^2& \lambda_{H\varphi} v_H v_\varphi
    \\ \lambda_{H\varphi} v_H v_\varphi  &  2 \lambda_\varphi v_\varphi^2\end{array}\right)
=
\left(\begin{array}{cc} m_1^2 c_H^2 + m_2^2 s_H^2& (m_2^2-m_1^2) c_H s_H \\ 
(m_2^2-m_1^2) c_H s_H  & m_1^2 s_H^2 + m_2^2 c_H^2\end{array}\right)
\label{eq:scalar_MM}
\eea
where $\alpha_H$ is obtained from the relation
\bea
\tan 2 \alpha_H =\frac{\lambda_{H\varphi} v_H v_\varphi}{\lambda_\varphi v_\varphi^2-\lambda_H v_H^2}.
\eea

There is a mass splitting between the real and imaginary part of $X$:
\bea
 X=\frac{X_R + i X_I}{\sqrt{2}}.
\eea
In the scalar potential we have 22 parameters in total. We can trade some of those parameters for masses,
\bea
\mu_X^2 &=& {1 \over 2} (m_R^2 + m_I^2 -\lambda_{HX} v_H^2 - \lambda_{\varphi X} v_\varphi^2),\\
\mu_{\varphi X} &=& \frac{m_R^2-m_I^2}{2 \sqrt{2} v_\varphi}, \label{eq:muphiX} \\
\mu_h^2 &=& m_{h^+}^2 -{1 \over 2} \lambda_{Hh} v_H^2 -{1 \over 2} \lambda_{\varphi h} v_\varphi^2,\\
\mu_k^2 &=& m_{k^{++}}^2 -{1 \over 2} \lambda_{Hk} v_H^2 -{1 \over 2} \lambda_{\varphi k} v_\varphi^2,
\eea
where $m_{R(I)}$ is the mass of $X_{R(I)}$.
For simplicity we take $X_R$ as the dark matter candidate from now on.
We can also express $\lambda_H, \lambda_\varphi, \lambda_{H\varphi}$ in terms of masses $m_i^2 (i=1,2)$ 
and mixing angle $\alpha_H$, then we take the 22 free parameters as
\bea
&& v_H (\simeq 246 \;{\rm GeV}), \quad v_\varphi, \quad m_1 (\simeq 125 \;{\rm GeV}), \quad m_2, \quad \alpha_H, \nl
&& m_R, \quad m_I, \quad m_{h^+}, \quad m_{k^{++}}, \nl
&&\lambda_\mu, \quad \lambda_h, \quad \lambda_k, \quad \lambda_X, \nl
&& \lambda_{Hh}, \quad \lambda_{Hk}, \quad \lambda_{HX}, \quad \lambda_{\varphi X}, 
\quad \lambda_{\varphi h}, \quad \lambda_{\varphi k}, \quad \lambda_{Xh}, \quad \lambda_{Xk}, 
\quad \lambda_{hk}, 
\eea
where two values, $v_H$ and $m_1$, have been measured as written in the parentheses.

\subsection{$X_R X_R \to \gamma \gamma$ and Fermi-LAT 130 GeV $\gamma$-ray excess in 
$U(1)_{B-L}$ model}

In this section we will see that we can obtain dark matter annihilation cross section into two photons, 
$X_R X_R \to \gamma \gamma$, large enough to explain the Fermi-LAT 130 GeV $\gamma$-line excess. There are two mechanisms to enhance the
annihilation cross section in this model: $H_2$-resonance and large $v_\varphi$.
In these cases, since the SM Higgs, $H_1$, contribution is small for small mixing angle $\alpha_H$, 
we consider only the contribution of $H_2$ assuming $\alpha_H=0$ (or $H_2 =\phi$).
Allowing nonvanishing $\al_H$ would only increase the allowed region of parameter space.
Then we obtain the annihilation cross section times relative velocity for $X_R X_R \to \gamma \gamma$,
\bea
\sigma v_{\rm rel} (X_R X_R \to \ga \ga)&=&\frac{\al_{\rm em}^2}{32\pi^3 s}
\left|\frac{(\sqrt{2} \mu_{\varphi X} + \la_{\varphi X} v_\varphi)v_\varphi}{s-m_\phi^2+i m_\phi \Ga_\phi}
\sum_{i=h,k} Q_i^2 \la_{\varphi i}  [1-\tau_i f(\tau_i)] \right. \nl
&& \left. +\sum_{i=h,k} Q_i^2 \la_{Xi}  [1-\tau_i f(\tau_i)] \right|^2,
\label{eq:XXrr}
\eea
where $Q_i$ is electric charge of $i(=h^+, k^{++})$, $\tau_i=4 m_i^2/s$ and $\Ga_\phi$ is total decay width of $\phi$. 
Since $v_{\rm rel} \approx 10^{-3} \ll 1$, we can approximate $s=4 m_R^2/(1-v_{\rm rel}^2/4) \approx 4 m_R^2$.
When $\al_H=0$, the $H_2(=\phi)$ can decay into two Goldstone bosons ($\alpha$) or into two photons
with partial decay width
\bea
\Ga(\phi \to \alpha\alpha) &=& \frac{m_\phi^3}{32 \pi v_\varphi^2},
\label{eq:phiGG} \\
\Ga(\phi \to X_{R(I)} X_{R(I)}) &=& \frac{(\pm \sqrt{2} \mu_{\varphi X}+\la_{\varphi X} v_\varphi)^2}{32 \pi m_\phi}
\sqrt{1-\frac{4 m_{R(I)}^2}{m_\phi^2}}, 
\label{eq:phiXX}\\
\Ga(\phi \to h^{+} h^{-} (k^{++} k^{--})) &=& \frac{(\la_{\varphi h(k)} v_\varphi)^2}{16 \pi m_\phi} 
\sqrt{1-\frac{4 m_{h^+(k^{++})}^2}{m_\phi^2}}, \\
\Ga(\phi \to \ga\ga) &=& \frac{\al_{\rm em}^2 v_\varphi^2}{64\pi^3 m_\phi}
\left|\sum_{i=h,k} Q_i^2 \la_{\varphi i}  [1-\tau_i f(\tau_i)] \right|^2.
\eea
Then the total decay width of $\phi$ is the sum:
\bea
\Ga_\phi =\Ga(\phi \to \alpha\alpha) +\Ga(\phi \to X_{R(I)} X_{R(I)})+\Ga(\phi \to h^{+} h^{-} (k^{++} k^{--}))+ \Ga(\phi \to \ga\ga).
\eea

\begin{figure}
\center
\includegraphics[width=7cm]{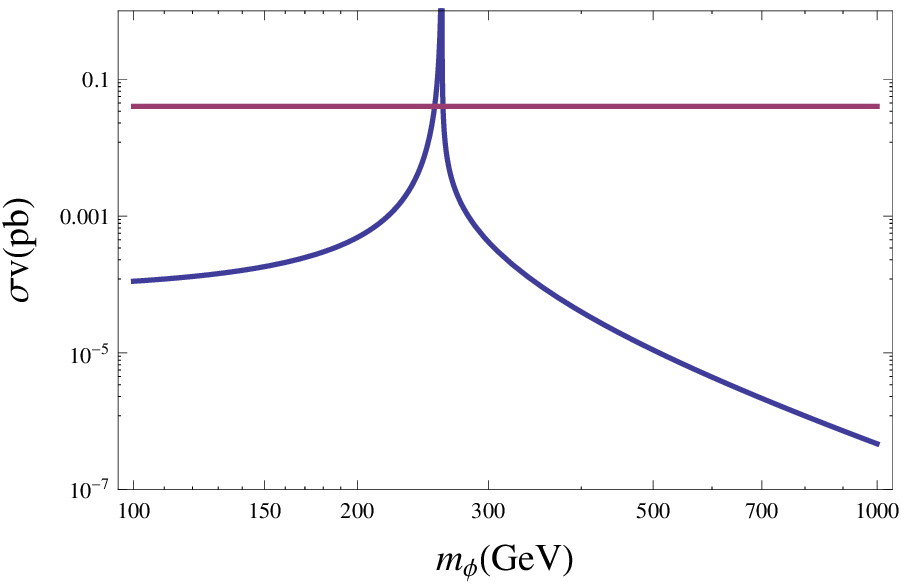}
\includegraphics[width=7cm]{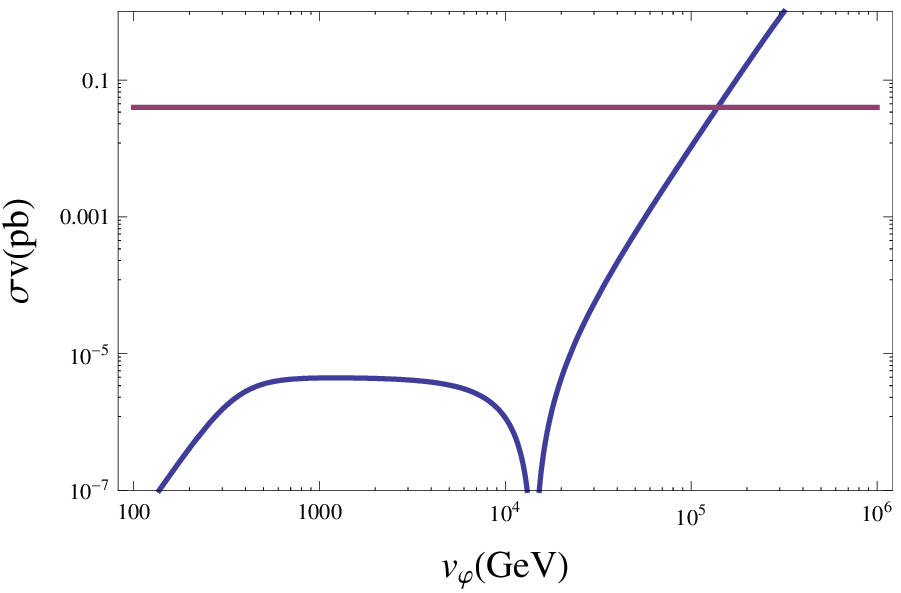}
\caption{Plots of $\si v (X_R X_R \to \ga \ga)$ for $\al_H=0$ as a function of $m_\phi (=m_2)$ (left panel) and
$v_\varphi$ (right panel). We set $m_R=130$, $m_I=2000$, $m_{h^+}=300$, $m_{k^{++}}=500$ (GeV), 
$\la_{\varphi h} = \la_{\varphi k} =0.1$, $\la_{\varphi X} = \la_{X h} = \la_{X k} =0.01$,  
$v_\varphi =1000$ (GeV) for the left panel
and $m_\phi=600$ (GeV) for the right panel. 
The horizontal red line represent $\si v(X_R X_R \to \ga\ga)=0.04$ (pb)
which can explain the Fermi-LAT gamma-line signal.
}
\label{fig:XXrr}
\end{figure}

As mentioned above, Fig.~\ref{fig:XXrr} shows the two enhancement mechanisms for $X_R X_R \to \ga \ga$:
the left panel for the $\phi$-resonance and the right panel for the large $v_\varphi$. For these plots we set the parameters:
$m_R=130$, $m_I=2000$, $m_{h^+}=300$, $m_{k^{++}}=500$ (GeV), 
$\la_{\varphi h} = \la_{\varphi k} = 0.1$, $\la_{\varphi X}= \la_{X h} = \la_{X k} =0.01$,  $v_\varphi =1000$ (GeV) for the left plot
and $m_\phi=600$ (GeV) for the right plot. We can obtain the large annihilation cross section required
to explain Fermi-LAT gamma-line data either near the resonance, $m_\phi \approx 2 m_R$ (left panel) or at large $v_\varphi$
(right panel). These behaviors can be understood easily from (\ref{eq:XXrr}). 
In either of these cases only the 1st term in (\ref{eq:XXrr}) gives large enhancement. 
The slope on the right of the resonance peak (the left panel of Fig.~\ref{fig:XXrr}) is steeper than that on
the left because, when $m_\phi >260$ GeV, new annihilation channel $\phi \to X_R X_R$ 
opens and the decay width of $\phi$
increases leading to decreasing the annihilation cross section.
In the right panel of Fig.~\ref{fig:XXrr}, the dip near $v_\varphi \approx 10^4$ GeV occurs because
there is cancellation between $\sqrt{2} \mu_{\varphi X} =(m_R^2 - m_I^2)/(2 v_\varphi)$ and $\la_{\varphi X} v_\varphi$
terms for positive $\la_{\varphi X}$.

\begin{figure}
\center
\includegraphics[width=9cm]{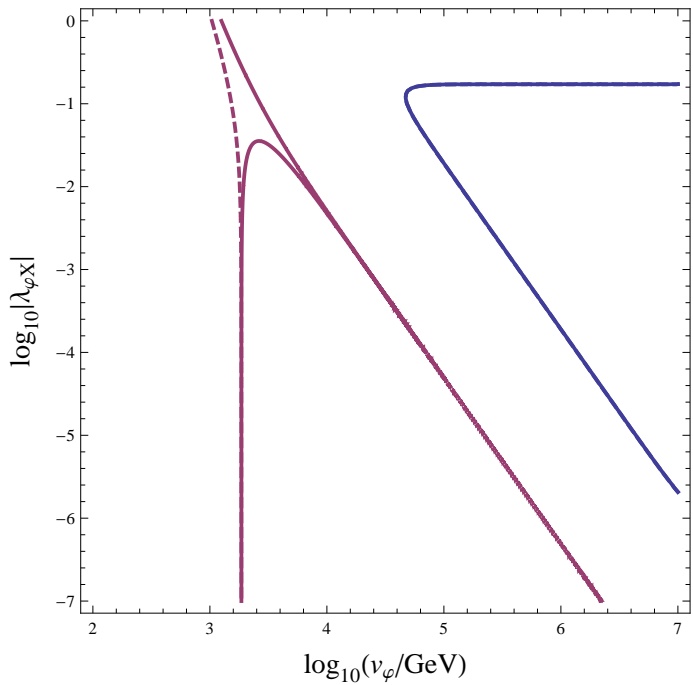}
\caption{Contour plot of $\si v (X_R X_R \to \ga \ga)=0.04$ (pb) in $(v_\varphi,\la_{\varphi X})$-plane.
The red lines represent the $\phi$-resonance solution and
the blue lines represent the large $v_\varphi$ solution. The solid (dashed) lines are for positive (negative)
$\la_{\varphi X}$. See the text for the parameters chosen for this plot.
}
\label{fig:XXrr_cont}
\end{figure}

The two mechanisms can also be seen in Fig.~\ref{fig:XXrr_cont}. This figure shows a contour plot of
$\si v (X_R X_R \to \ga \ga)=0.04$ (pb) in $(v_\varphi,\la_{\varphi X})$-plane.
 We set $m_R=130$, $m_I=1000$, $m_{h^+}=1000$, $m_{k^{++}}=1000$,  $m_\phi =260$ (GeV), 
$\la_{\varphi h} = \la_{\varphi k} = \la_{X h} = \la_{X k} =0.01$ for red lines ($\phi$-resonance).
And we take $m_I=1000$, $m_{h^+}=300$, $m_{k^{++}}=500$,  $m_\phi =600$ (GeV),
$\la_{\varphi h} = \la_{\varphi k} =0.1$,  $\la_{X h} = \la_{X k} =0.01$ for blue line (large $v_\varphi$). 
The red (blue) lines represent the $\phi$-resonance (large $v_\varphi$) solution for Fermi-LAT anomaly.
In the $\phi$-resonance region, for the  negative (positive) $\la_{\varphi X}$ 
the two values $\sqrt{2} \mu_{\varphi X}=(m_R^2-m_I^2)/(2 v_\varphi)$ and $\la_{\varphi X} v_{\varphi}$
which appear in the 1st term of (\ref{eq:XXrr}) have the same (opposite) sign and
their contributions are constructive (destructive). As a result for positive $\la_{\varphi X}$ (solid red line),
there is cancellation between the two terms, and  larger value
of $v_\varphi$ is required for a given $\la_{\varphi X}$. 
For large $v_\varphi$ case, the result does not depend on the sign of $\la_{\varphi X}$ because the 
$\la_{\varphi X} v_{\varphi}$ term dominates. And the solid and dashed blue lines overlap each other in 
Fig.~\ref{fig:XXrr_cont}. For $\la_{\varphi X}$ larger than about $0.1$ the decay width 
$\Ga(\phi \to X_R X_R)$ becomes too large to enhance the annihilation cross section.

\subsection{Relic density in $U(1)_{B-L}$ model}
Now we need to check whether the large enhancement in $X_R X_R \to \ga\ga$ signal is consistent with
the observed relic density $\Om_{\rm DM} h^2 =0.1199 \pm 0.0027$.
To obtain the current relic density the DM annihilation cross section at decoupling time should be approximately
(assuming $S$-wave annihilation)
\bea
\langle \si v \rangle_{\rm th} \approx 3 \times 10^{-26} \;{\rm cm^3/s} \approx 1 \;{\rm pb},
\eea
from (\ref{eq:Om}).
The major difference between the $Z_2$ model and the $U(1)_{B-L}$ model is that the latter model has additional
annihilation channel, {i.e.}, $X_R X_R \to \al \al$ and $\phi$-exchange $s$-channel diagrams
compared with the former one. The S-wave contribution to $\si v(X_R X_R \to \al \al)$ is
shown in the Appendix.
The Goldstone boson mode becomes dominant especially when $v_\varphi$ is not very 
large~\cite{Lindner:2011it}, {\it i.e.} $v_\varphi \lesssim 10^3$ GeV. And it makes the dark matter phenomenology very different from the one
without it. For example, in $Z_2$ model we need the annihilation channel $X X \to h^+ h^- (k^{++} k^{--})$
large enough to obtain the current relic density. In $U(1)_{B-L}$ model, however, the annihilation
into Goldstone bosons are sometimes large enough to explain the relic density.\footnote{The dark sector
can be in thermal equilibrium with the SM plasma in the early universe even with very small
mixing $\al_H \sim 10^{-8}$~\cite{Baek:2014goa}. And  our analysis with $\al_H =0 $ can be thought of
as a good approximation of more realistic case of non-zero but small $\al_H$.}

\begin{figure}
\center
\includegraphics[width=8cm]{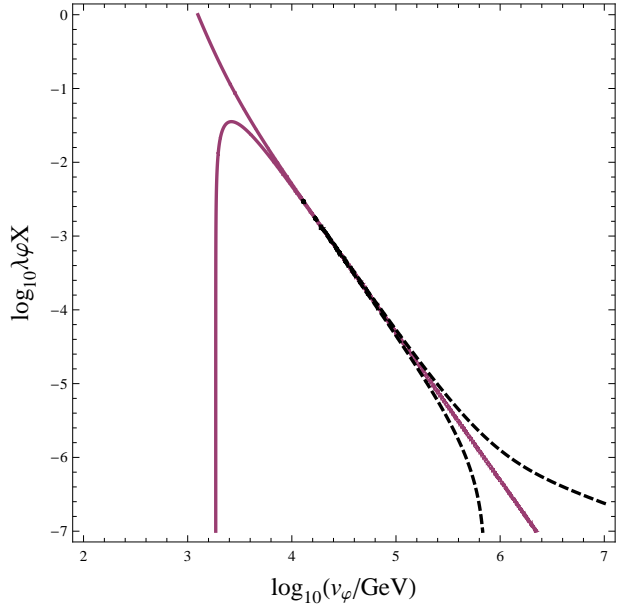}
\caption{Contour plots of $\si v (X_R X_R \to \ga \ga)=0.04$ pb (solid red line) and 
$\Om_{DM} h^2 =0.1199$ (dashed black line) in the $(v_\varphi, \la_{\varphi X})$-plane for $\la_{\varphi X}>0$.
We take the parameters, $2 m_R = m_\phi =260$ GeV. See the text for other parameters. The region
enclosed by the dashed lines gives $\Om_{\rm DM} h^2 >0.12$.}
\label{fig:pos}
\end{figure}

To see the relevant parameter space satisfying both the Fermi-LAT 130 GeV gamma-line anomaly
and the correct relic density, we consider the $\phi$-resonance and large $v_\varphi$ cases discussed above
separately. Fig.~\ref{fig:pos} shows contours of $\si v(X_R X_R \to \ga\ga)=0.04$ (pb) (solid line) 
and $\Om_{X_R} h^2 \approx 0.12$ (dashed line)
for $\la_{\varphi X} >0$ when the resonance condition $m_\phi = 2 m_R$ is satisfied. 
The parameters are chosen as $m_\phi=2 m_R = 260$ GeV, $m_I=m_{h^+}=m_{k^{++}}=1$ TeV,
$\la_{\varphi h}= \la_{\varphi k}= \la_{X h}= \la_{X k}=0.01$. We can see there are intersection points of
the two lines where both Fermi-LAT anomaly and the relic density can be explained. 
For the parameters we have chosen  the contribution of $X_R X_R \to \al\al$ to the relic density
is almost 100\%. This implies there is wide region of allowed parameter space satisfying
both observables, since other annihilation channels $X X \to h^+ h^- (k^{++} k^{--})$ are also 
available when they are kinematically allowed.
Typically TeV scalar $v_\varphi$ gives too large $X_R X_R \to \al\al$ annihilation cross section resulting
in too small relic density. For the positive $\la_{\varphi X} $ case, however, there is also cancellation
between terms in $\si v(X_R X_R \to \al\al)$ as in $\si v(X_R X_R \to \ga\ga)$. Both cancellations are
effective when the condition, $\la_{\varphi X} v_\varphi^2 =(m_I^2 -m_R^2)/2$, is satisfied. 
This explains the intersection point occurs on the diagonal straight line determined by the above condition.
This allows large relic density even near TeV $v_\varphi$.

\begin{figure}
\center
\includegraphics[width=8cm]{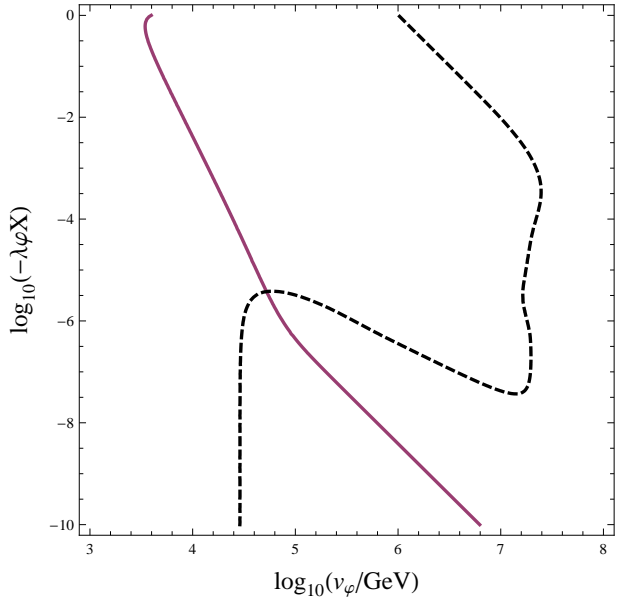}
\caption{The same plot with Fig.~\ref{fig:pos} for $\la_{\varphi X}<0$.
We also take the $\phi$-resonance condition, $2 m_R = m_\phi =260$ GeV. See the text for other parameters.
The region to the right of the dashed line gives $\Om_{\rm DM} h^2 >0.12$.}
\label{fig:neg}
\end{figure}

\begin{figure}
\center
\includegraphics[width=8cm]{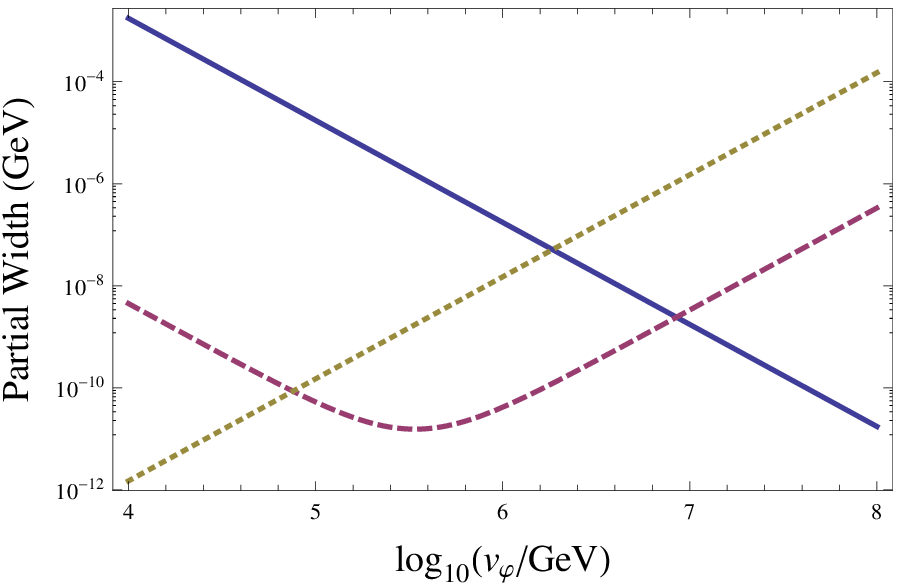}
\caption{Plots of $\Ga(\phi \to \al\al)$ (solid blue line), $\Ga(\phi \to X_R X_R)$ (dashed red line)
and $\Ga(\phi \to \ga \ga)$ (dotted green line)  as a function of $v_\varphi$ for the
parameters used in Fig.~\ref{fig:neg}. We fixed $\la_{\varphi X}=-10^{-7}$.}
\label{fig:width}
\end{figure}

Fig.~\ref{fig:neg} shows the same contours for $\la_{\varphi X} <0$. In this case as we have seen in 
Fig.~\ref{fig:XXrr_cont} that TeV scale $v_\varphi$ can explain Fermi-LAT gamma-line. However this
value of $v_\varphi$ gives too large DM annihilation cross section at the decoupling time (when
$X_R X_R \to \al\al$ is dominant) and too small relic density. So somehow we need to ``decouple''
the $X_R X_R \to \ga\ga$ so that we need larger $v_\varphi$. We can do it, for example, by assuming $h^+ (k^{++})$
are very heavy: $m_{h^+} =m_{k^{++}} =5$ TeV.  If we also reduce the mass difference $m_I -m_R$, we get
smaller $\sqrt{2} \mu_{\varphi X}=-(m_I^2-m_R^2)/(2 v_\varphi)$.  Then we have simultaneous solution both
for $\si v(X_R X_R \to \ga\ga) \approx 0.04$ pb and $\Om_{X_R} h^2 \approx 0.12$ as can be seen in Fig.~\ref{fig:neg}.
The two lines meet at rather large $v_\varphi(\sim 10^5 \; {\rm GeV})$ as expected.
For other parameters we chose\footnote{To avoid fine tuning
we allowed small off-resonance condition of the size of $\Ga_\phi\sim 1$ keV, {\it i.e.},
we set $m_\phi=260.000001$ GeV.} $m_\phi\approx 2 m_R = 260$ GeV, $m_I=200$ GeV,
$\la_{\varphi h}= \la_{\varphi k}= \la_{X h}= \la_{X k}=0.01$. The pattern of the relic density contour requires some explanation.
The annihilation cross section for $X_R X_R \to \al \al$ at the resonance can be
approximated from (\ref{eq:XXGG}) as
\bea
\si v(X_R X_R \to \al \al) \approx \frac{(\sqrt{2} \mu_{\varphi X} + \la_{\varphi X} v_\varphi)^2}{16\pi v_\varphi^2 \Ga_\phi^2},
\label{eq:ann_approx}
\eea
where $\Ga_\phi$ is the total decay width of $\phi$.
The nonvanishing partial decay widths of $\phi$ for the parameters we chose are
$\Ga(\phi \to \al\al)$, $\Ga(\phi \to X_R X_R)$ and $\Ga(\phi \to \ga\ga)$. 
In Fig.~\ref{fig:width} they are plotted as a function of
$v_\varphi$ for $\la_{\varphi X} =-10^{-7}$. On the vertical part of the relic density contour in Fig.~\ref{fig:neg} near $v_\varphi \sim 10^{4.4}$ GeV,
the $\Ga(\phi \to\al\al)$ dominates and also $\sqrt{2} \mu_{\varphi X} \gg \la_{\varphi X} v_\varphi$. For this $v_\varphi$
we approximately get
\bea
\si v(X_R X_R \to \al \al) \sim \frac{16\pi m_I^4}{m_\phi^6},
\eea
which is independent of $\la_{\varphi X}$.
Around $v_\varphi \sim 10^{7.2}$ GeV, 
$\Ga(\phi \to X_R X_R)$ and $\Ga(\phi \to \ga\ga)$ dominate despite high phase space suppression
in $\phi \to X_R X_R$, and $\sqrt{2} \mu_{\varphi X} \ll \la_{\varphi X} v_\varphi$.
As $\la_{\varphi X}$ increases, $\Ga(\phi \to X_R X_R)$ becomes more important than  $\Ga(\phi \to \ga\ga)$
as can be seen from (\ref{eq:phiGG}) and (\ref{eq:phiXX}). 
The (almost) vertical part for this $v_\varphi$ region is due to partial cancellation of the factor
$(\sqrt{2} \mu_{\varphi X} + \la_{\varphi X} v_\varphi)^2$ in the numerator of (\ref{eq:ann_approx}) and
the same factor in the cross term of $\Ga(\phi \to X_R X_R)$ and $\Ga(\phi \to \ga\ga)$
in the denominator. As $\la_{\varphi X}$ grows even larger, only $\Ga(\phi \to X_R X_R)$ term dominates and
$\si v(X_R X_R \to \al \al) \propto 1/\la_{\varphi X}^2 v_\varphi^4$, which gives the slanted part of the contour line.

\begin{figure}
\center
\includegraphics[width=8cm]{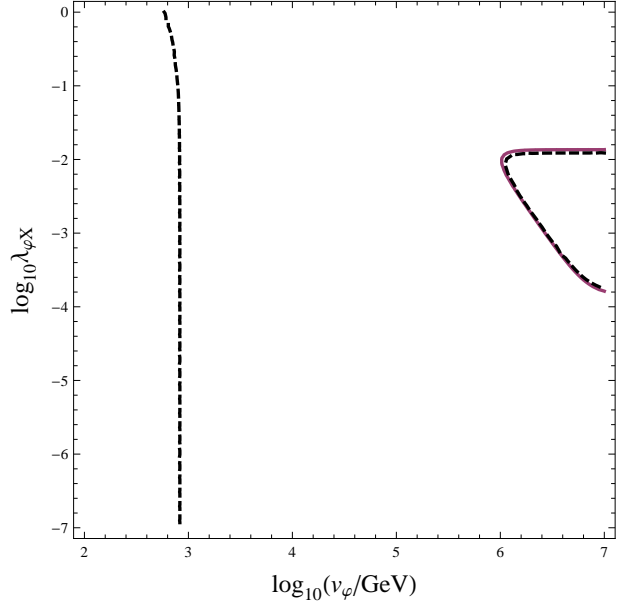}
\caption{The same plot with Fig.~\ref{fig:pos} corresponding to large $v_\varphi$ solution.
See the text for the parameters used in this figure.}
\label{fig:off}
\end{figure}
\begin{figure}
\center
\includegraphics[width=8cm]{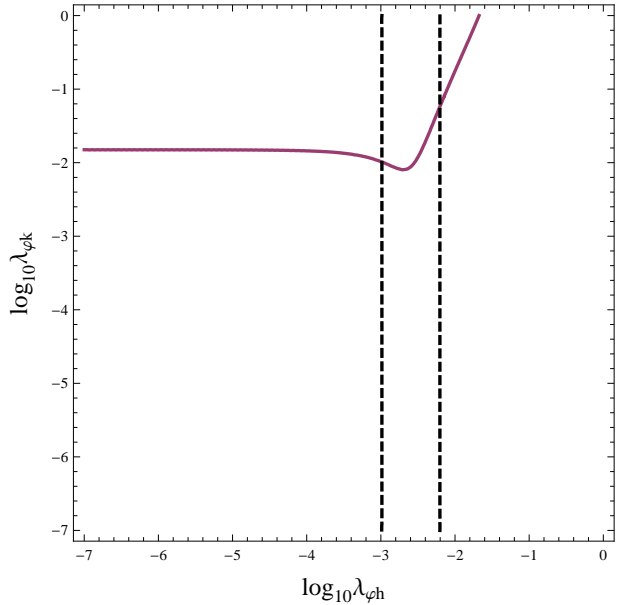}
\caption{Contours of $\sigma v(X_R X_R \to \ga\ga)=0.04$ pb (solid red line)
and $\Om h^2=0.1199$ (dashed blue line).}
\label{fig:off_la}
\end{figure}

We can also obtain simultaneous solutions when $\phi$ is off-resonance using large $v_\varphi$. 
Fig.~\ref{fig:off} shows an example of this case. In this case, if we have only $X_R X_R \to \al\al$-channel for relic
density, the resulting $\Om_{X_R} h^2$ is too large for $v_\varphi \gtrsim 1$ TeV. To get the correct
relic density by increasing the DM pair annihilation cross section at freeze-out
we allowed $X_R X_R \to h^+ h^-$ channel. Then we can get a solution as can be
seen in Fig.~\ref{fig:off}. The region enclosed by two dashed lines overcloses the universe.
For this plot, we chose $m_R=130$ GeV, $m_\phi=m_I=1$ TeV, 
$m_{h^+} =150$  GeV, $m_{k^{++}}=500$ GeV, $\la_{\varphi h}=0.001$, and $\la_{\varphi h}= \la_{\varphi k}= \la_{X h}= \la_{X k}=0.01$. 
Note that we take $\la_{\varphi h}=0.001$ so that the solid red line representing $\sigma v(X_R X_R \to \ga\ga)=0.04$ pb
and dashed line representing $\Om h^2=0.1199$ overlap with each other.  To show that this choice of $\la$'s
is possible in general, we take a point on the overlapped lines, {\it e.g.}, $v_\varphi = 10^{6.43}$ GeV and 
$\la_{\varphi X}=0.001$. Then we can get $\la_{\varphi h}=0.001, \la_{\varphi h}=0.01$ as a solution from Fig.~\ref{fig:off_la}.

When there is no mixing between $\phi$ and $h$ the decay width $H_1 \to \ga\ga$ is the same
with that of the SM. This means that we can enhance the $X_R X_R \to \ga\ga$ without affecting
the SM $H_1 \to \ga\ga$ rate. When the mixing angle $\al_H$ is nonvanishing the
$h^+$ and/or $k^{++}$ can contribute to $H_1 \to \ga\ga$ through one-loop process.
Since this effect was already discussed in Sec.~\ref{sec:Hrr}, we do not discuss it further.

Weinberg~\cite{Weinberg:2013kea} showed that Goldstone bosons can play the role of dark radiation
and contribute to the effective number of neutrinos $N_{\rm eff}$. When Goldstone bosons go out of
equilibrium when the temperature is above the mass of muons and electrons but below that of all other
SM particles, we get $\De N_{\rm eff} =0.39$. The condition for this to happen is~\cite{Weinberg:2013kea}
\bea
\frac{\la_{H\varphi}^2 m_\mu^7 M_{\rm pl}}{m_\phi^4 m_H^4} \approx 1.
\eea
For example, with $\lambda_{H\varphi}=0.005$ and $m_H = 125$ GeV, the dark scalar with mass 500 MeV can
satisfy the condition.

\section{Conclusions}
\label{sec:Conclusions}
We have considered two scenarios which minimally extended the `Zee-Babu model'~\cite{Zee,Babu,Babu:2003}.
In the first scenario we introduced a real scalar dark matter  $X$ with $Z_2$ symmetry:  
$X \rightarrow -X$. If the scalar dark matter $X$ has a mass around 130 GeV, 
the annihilation cross section, $\langle \sigma v \rangle(XX \to \gamma\gamma)$, can be enhanced by
the contribution of the singly- and/or doubly-charged Zee-Babu scalars.
If we also want to explain the dark matter relic abundance, however, we get at most 
$\langle \sigma v \rangle_{\gamma\gamma} \approx 0.2 \times 10^{-27} {\rm cm^3/s} $, which
is about factor 6 smaller than the required value to explain the Fermi-LAT gamma-ray line signal.

We have shown that the present constraint on the couplings $\lambda_{Xk}$
and $\lambda_{Xh}$ which mix the dark matter and charged Higgs 
is not so strong and they can enhance the annihilation
cross section of $X X \to \gamma \gamma$ large enough to accommodate
the recent hint.
On the other hand the couplings which involve the SM Higgs $H$
are strongly constrained by the theoretical considerations in the
Higgs potential and the observations of dark matter relic density
and dark matter direct detections.
The upper bound on the $\lambda_{HX}$ coupling is about 0.06 which
comes from the dark matter direct detection experiments.
For the $\lambda_{Hh}, \lambda_{Hk}$ which mix the SM Higgs and the
new charged Higgs, the theoretical bound becomes more important.
If we require the absolute stability of the dark matter by the
$Z_2$ symmetry $X \to -X$ and the absence of charge breaking, we get
the upper bound of $\lambda_{Hh},\lambda_{Hk}$ to be about 0.7 for the charged Higgs mass around 150 GeV.
To evade the unbounded-from-below Higgs potential we need to have
$\lambda_{Hh}, \lambda_{Hk} \gtrsim -1.6$.
With these constraints the $B(H \to \gamma (Z) \gamma)$ can be enhanced up to
1.5 (1.1) or suppressed down to 0.5 (0.9) with respect to that in the SM.
%
The neutrino sector cannot be described by the Zee-Babu model only, 
and there should be additional contributions to the neutrino masses and mixings
such as dimension-5 Weinberg operator from type-I seesaw mechanism.

In the second scenario, we introduced two complex scalar fields $X$ and $\varphi$ 
with global $U(1)_{B-L}$ symmetry. After $\varphi$ gets vev, $v_\varphi$, the $U(1)_{B-L}$ symmetry is broken down
to $Z_2$ symmetry. The lighter component of $X$, which we take to be the real part, $X_R$, is
stable due to the remnant $Z_2$ symmetry and can be a dark matter candidate.
Even in the extreme case where we do not consider the mixing of the dark scalar and the standard model
Higgs scalar ($\alpha_H=0$), we showed that the dark matter relic abundance and the Fermi-LAT gamma-ray
line signal can be accommodated in two parameter regions: resonance region ($m_\phi=2 m_R$) and
large $v_\varphi(\sim 10^6 -10^7 {\rm GeV})$ region. Since there is no mixing, there is no
correlation with $H \to \gamma\gamma$ and direct detection scattering of dark matter off the proton.
In addition the neutrino sector need not be modified contrary to the first scenario.

\acknowledgments
This work is partly supported by NRF Research Grant  2012R1A2A1A01006053 (PK, SB).

\appendix
\section{One-loop $\beta$ functions of the quartic couplings}\label{app:RGE}
Here, we give the renormalization group equation and the one-loop $\beta$ functions of the quartic couplings:
\begin{align}
\frac{d\lambda_i}{d\ln Q}=\beta_{\lambda_i},
\end{align}
with
\begin{align}
\beta_{\lambda_H} &=\frac{1}{16\pi^2}
\bigg[
24\lambda_H^2+\lambda_{Hh}^2+\lambda_{Hk}^2+\frac{\lambda_{HX}^2}{2}-6y_t^4
+\frac{3}{8}\Big\{2g_2^4+(g_2^2+g_1^2)^2\Big\} \nonumber\\
&\hspace{8cm} -4\lambda_H\left\{\frac{3}{4}(3g_2^2+g_1^2)-3y_t^2\right\}
\bigg], \\
\beta_{\lambda_h} &=\frac{1}{16\pi^2}
\left[
16\lambda_h^2+2\lambda_{Hh}^2+\lambda_{hk}^2+\frac{\lambda_{Xh}^2}{2}+6g_1^4
	-12\lambda_hg_1^2
\right], \\
\beta_{\lambda_k} &=\frac{1}{16\pi^2}
\left[
16\lambda_k^2+2\lambda_{Hk}^2+\lambda_{hk}^2+\frac{\lambda_{Xk}^2}{2}+96g_1^4
	-48\lambda_kg_1^2
\right],\\
\beta_{\lambda_X} &= \frac{1}{16\pi^2}
\Big[
18\lambda_X^2+2\lambda_{HX}^2+\lambda_{Xh}^2+\lambda_{Xk}^2
\Big], \\
\beta_{\lambda_{Hh}} &=\frac{1}{16\pi^2}
\bigg[
12\lambda_H\lambda_{Hh}+8\lambda_h\lambda_{Hh}+2\lambda_{hk}\lambda_{Hk}
+\lambda_{HX}\lambda_{Xh}+3g_1^4 \nonumber\\
&\hspace{6cm}
	-\lambda_{Hh}\bigg\{\frac{3}{2}(3g_2^2+g_1^2)-6y_t^2+6g_1^2\bigg\}
\bigg], \\
\beta_{\lambda_{Hk}} &=\frac{1}{16\pi^2}
\bigg[
12\lambda_H\lambda_{Hk}+8\lambda_k\lambda_{Hk}+2\lambda_{hk}\lambda_{Hh}
+\lambda_{HX}\lambda_{Xk}+12g_1^4 \nonumber\\
&\hspace{6cm}
	-\lambda_{Hk}\bigg\{\frac{3}{2}(3g_2^2+g_1^2)-6y_t^2+24g_1^2\bigg\}
\bigg], \\
\beta_{\lambda_{hk}} &=\frac{1}{16\pi^2}
\Big[
4\lambda_{Hh}\lambda_{Hk}+8\lambda_{hk}(\lambda_h+\lambda_k)+\lambda_{Xh}\lambda_{Xk}
	+48g_1^4-30\lambda_{hk}g_1^2
\Big],\\
\beta_{\lambda_{HX}} &= \frac{1}{16\pi^2}
\bigg[
	12\lambda_H\lambda_{HX}+2\lambda_{Hh}\lambda_{Xh}+2\lambda_{Hk}\lambda_{Xk}
	+6\lambda_X\lambda_{HX}
	-\lambda_{HX}\bigg\{\frac{3}{2}(3g_2^2+g_1^2)-6y_t^2\bigg\}
\bigg], \\
\beta_{\lambda_{Xh}} &= \frac{1}{16\pi^2}
\Big[
4\lambda_{Hh}\lambda_{HX}+8\lambda_h\lambda_{Xh}+2\lambda_{hk}\lambda_{Xk}
+6\lambda_X\lambda_{Xh}-6\lambda_{Xh}g_1^2
\Big], \\
\beta_{\lambda_{Xk}} &= \frac{1}{16\pi^2}
\Big[
4\lambda_{Hk}\lambda_{HX}+8\lambda_k\lambda_{Xk}+2\lambda_{hk}\lambda_{Xh}
+6\lambda_X\lambda_{Xk}-24\lambda_{Xk}g_1^2
\Big].
\end{align}

\section{The  annihilation cross section of $X_R X_R \to \al \al$}
The  annihilation cross section of $X_R X_R \to \al \al$ is obtained as:
\bea
\si v(X_R X_R \to \al \al)&=&\frac{m_R^2 \Big[4 v_\varphi (\sqrt{2} \mu_{\varphi X} +\la_{\varphi X} v_\varphi)(m_I^2+m_R^2)+(m_I^2-m_R^2)(4
  m_R^2-m_\phi^2)\Big]^2}{64 \pi v_\varphi^4 (m_I^2+m_R^2)^2(4 m_R^2-m_\phi^2)^2} \nl
&& + O(v^2),
\label{eq:XXGG}
\eea
where $(4 m_R^2-m_\phi^2)^2$ should be replaced by $(4 m_R^2-m_\phi^2)^2+ \Ga_\phi^2 m_\phi^2$ in the
denominator when $2 m_R \approx m_\phi$.


\end{document}